 \documentclass[final,3p,times,authoryear]{elsarticle}
\usepackage{amssymb}
\usepackage{CJKutf8}
\usepackage{float}
\usepackage{multirow}
\usepackage{color, colortbl}
\usepackage{graphicx}

\usepackage{amsmath}
\DeclareMathOperator*{\argmin}{arg\,min}
\DeclareMathOperator*{\argmax}{arg\,max}

\usepackage{natbib}
\usepackage[pdftex,colorlinks,linkcolor=blue,citecolor=blue,urlcolor=blue,
bookmarks=true,bookmarksopen=false,bookmarksnumbered=true]{hyperref}

\newcommand{\comment}[1]{}

\newcommand{\tabincell}[2]{\begin{tabular}{@{}#1@{}}#2\end{tabular}}

\makeatletter
\AtBeginDvi{\pdfmapline{=simkai@Unicode@  < simkai-select.ttf}}
\DeclareFontFamily{C70}{simkai}{\hyphenchar \font\m@ne}
\DeclareFontShape{C70}{simkai}{l}{n}{ <-> CJK * simkai}{}
\DeclareFontShape{C70}{simkai}{m}{n}{ <-> CJK * simkai}{\CJKnormal}
\DeclareFontShape{C70}{simkai}{bx}{n}{ <-> CJKb * simkai}{\CJKbold}
\makeatother

\begin{document}
\begin{CJK*}{UTF8}{simkai}

\begin{frontmatter}




\title{An Information Retrieval Approach to Short Text Conversation\tnoteref{mytitlenote}}
\tnotetext[mytitlenote]{This paper is an extended version of \citet{Wang:2013:dataset_stc} with the following new content: (1) empirical verification of effectiveness of several new matching models including translation model and deep matching model, and (2) proposal of topic-word model.}

\author[BJ]{Zongcheng Ji}
\corref{mycorrespondingauthor}
\cortext[mycorrespondingauthor]{Corresponding author.}
\ead{jizongcheng@gmail.com}

\author[HK]{Zhengdong Lu}
\ead{lu.zhengdong@huawei.com}

\author[HK]{Hang Li}
\ead{hangli.hl@huawei.com}

\address[BJ]{Noah's Ark Lab, Huawei Technologies, Beijing, China}
\address[HK]{Noah's Ark Lab, Huawei Technologies, Hong Kong, China}

\begin{abstract}

Human computer conversation is regarded as one of the most difficult problems in artificial intelligence. In this paper, we address one of its key sub-problems, referred to as short text conversation, in which given a message from human, the computer returns a reasonable response to the message. We leverage the vast amount of short conversation data available on social media to study the issue. We propose formalizing short text conversation as a search problem at the first step, and employing state-of-the-art information retrieval (IR) techniques to carry out the task. We investigate the significance as well as the limitation of the IR approach. Our experiments demonstrate that the retrieval-based model can make the system behave rather ``intelligently'', when combined with a huge repository of conversation data from social media.

\end{abstract}


\end{frontmatter}


\section{Introduction}
\label{sec:introduction}

\comment{Background}
Human computer conversation is one of the most challenging AI problems, which involves language understanding, reasoning, and use of common sense knowledge. Despite a significant amount of effort on the research in the past decades, the progress on the problem is unfortunately quite limited. One of the major reasons for that is lack of large volumes of real conversation data \citep{Chen:2011:QA_Generation, Nouri:2011:QA_Augmenting}.

\comment{Definition of STC}
In this paper, we consider a much simplified version of the problem: one round of conversation formed by two short texts, with the former being a message from human and the latter being a response to the message from the computer. We refer to it as \textbf{short text conversation (STC)}. Thanks to the extremely large amount of short text conversation data available on social media such as Twitter\footnote{http://twitter.com/} and Weibo\footnote{http://weibo.com/}, we anticipate that significant progress could be made in the research on the problem with the use of the big data, much like what has happened in machine translation, community question answering, etc.

\comment{Value of STC}
Modeling a short text conversation is much simpler than modeling a complete dialogue, which often requires several rounds of interactions (e.g., a dialogue system as in \citet{Litman:2000:Njfun}). However, it can shed important light on understanding of the complicated mechanism of natural language dialogues and can significantly enhance the research toward the ultimate goal of passing the Turing test.
The research on the problem will instantly help applications such as chatbot at a web site, automatic short-message reply on mobile phone, and voice assistant like Siri\footnote{http://en.wikipedia.org/wiki/Siri}.
With the emergence of social media, as well as the spread of mobile devices, conversation via short texts has become an important way of communication for people in our time.

\comment{Definition of Retrieval-based STC}
One simple approach to STC, and perhaps the first approach which one would want to try, is to take it as an information retrieval (IR) problem, maintain a large repository of short text conversation data, and develop a conversation system mainly based on IR technologies. Given a message, the system retrieves related responses from the repository and returns the most reasonable response. That is to say, we would not generate a new response, but select the most suitable response (originally made to other messages) as reply to the current message. We refer to the former approach as \textbf{generation-based STC} and the latter approach as \textbf{retrieval-based STC}. With advanced IR technologies and a dataset with previously unthinkable volume, we would expect that the conversation system can behave almost like a human in each round of conversation.

\comment{STC vs. CQA}
Retrieval-based STC is similar to some IR tasks such as community question answering (CQA). In the former task, each instance consists of a message-response pair (or a post-comment pair in social media), while in the latter task, each instance consists of a question-answer pair. The differences are also evident, however. The messages tend to be longer than the responses in STC, while the answers tend to be longer than the questions in CQA. More importantly, the relations between texts are different in the two problems. The answers in CQA must be solutions to the questions, which mostly involves knowledge, while the responses in STC need only be explanations, opinions, or criticisms on the messages, which is more about appropriateness or human-likeness.

\comment{Content / Contributions of this paper}
In this paper, we try to answer the question of to what extent the IR approach can effectively manipulate STC. We first propose a framework, which employs a learning to rank method for training of the ranking model and matching models as features of the ranking model. The matching models include (1) basic matching models based on cosine similarities, (2) translation model, (3) latent space model (linear model), (4) deep matching model (non-linear), and (5) topic-word model. The latent space model and deep matching model are techniques recently developed for search and question answering, and the topic-word model is devised for STC in this paper. We then conduct large scale experiments with a dataset from Weibo, which we have created and released for research on STC. Our experiments show that the retrieval-based approach can achieve fairly good performance with the precision at position one being $0.64$. Experiments also show that all the matching models can significantly improve the performance. We also conduct case study to show the significance and limitation of the IR approach.

The contributions of this paper, which is an extension of our previous paper \citep{Wang:2013:dataset_stc}, include (1) proposal of the IR approach to STC, (2) empirical verification of effectiveness and restriction of the approach, (3) proposal of topic-word model for STC, and (4) development of public dataset for the research.

\comment{Organization of this paper}
The rest of this paper is organized as follows. Section \ref{sec:related_work} reviews the related work. Section \ref{sec:conversation_example} shows an example of conversation on Weibo. Section \ref{sec:framework} gives the definition of retrieval-based STC and a three-stage retrieval based framework to perform STC. Section \ref{sec:dataset} and section \ref{sec:matching_features} give details on the dataset and the matching features used in the framework. Section \ref{sec:experiment} and \ref{sec:case_study} describe the experimental results and case studies, respectively. Finally, the work is concluded and future research directions are identified in Section \ref{sec:conclusions}.

\section{Related Work}
\label{sec:related_work}

\subsection{Short Text Conversation}

\comment{rule-based or learning-based, no/less data}
Early work on modeling dialogues is either rule-based \citep{Weizenbaum:1966:Eliza} or learning-based \citep{Litman:2000:Njfun, Schatzmann:2006:Survey, Williams:2007:Partially}. These approaches require no data (e.g., rule based) or little data (e.g., reinforcement learning based) for training, but much manual effort in building the model, which is usually very costly. Furthermore, the coverage of the systems is also not satisfactory.

\comment{NPCEditor, less data}
An alternative approach is to build a dialogue system with a knowledge base consisting of large number of question-answer pairs. For example, the system in \citet{Leuski:2006:NPCEditor} and \citet{Leuski:2011:NPCEditor} selects the most suitable response to the current message from the question-answer pairs using a statistical language model in cross-lingual information retrieval. The major bottleneck of this approach is creation of the knowledge  base (i.e., question-answer pairs). Instead of building knowledge base by hand, \citet{Chen:2011:QA_Generation} and \citet{Nouri:2011:QA_Augmenting} propose augmenting a knowledge base with question-answer pairs derived from texts using a question generation tool. The results show that the augmented system can answer questions about new topics, with certain performance drop on questions about existing topics. The number of question-answer pairs obtained in this way is still small (a few thousands). Furthermore, the statistical language model can only match questions and answers at the word level, not at the semantic level, which may hinder the performance of the system.

\comment{big data is coming}
All the above systems work at small scale in the sense that they can only respond to a small variety of messages. Recently, with the fast development of social media, such as community question answering and microblog services, a very large amount of conversation data becomes available.
\citet{Ritter:2011:resp_gen} investigate the feasibility of conducting short text conversation by using statistical machine translation (SMT) techniques, as well as millions of naturally occurring conversation data in Twitter. The results show that phrase-based SMT \citep{Koehn:2007:Moses} works better than vector space model (VSM) \citep{Salton:1975:VSM} in IR in terms of BLEU score \citep{Papineni:2002:BLEU}.  In the approach, a response is generated from a model, not retrieved from a repository, and thus it cannot be guaranteed to be a legitimate natural language text.

In this paper, we conduct short text conversation (STC) by leveraging the state-of-the-art IR technologies and the vast amount of conversation data on social media. A related but slightly different problem has been studied in \citet{Jafarpour:2010:Learning2Chat}, as an initial step for building a chatbot, referred to as learning to chat (L2C). L2C attempts to perform human computer conversation by utilizing machine learning and large scale dialogue data (each instance consists of several rounds of conversation). Their method first filters and then ranks responses to find the best candidate, on the basis of all information in the context. It is, thus, not directly comparable with our method in this paper.

\subsection{Search}

Learning to rank and semantic matching are considered state-of-the-art techniques for search \citep{Liu:2009:L2R, Li:2011:L2R, Li:2011:ShortL2R, Li:2014:SemanticMatch}. Given a query, documents containing the query terms are first retrieved from the index. Matching between the query and each of the documents is then carried out using different models such as traditional IR model of BM25 \citep{Robertson:1995:BM25}, translation model \citep{Berger:1999:Trans4IR}, and latent space model \citep{Wu:2013:bilinear}. The matching scores of each document are taken as features of the document. Next, the documents are assigned scores by the ranking model on the basis of their features. Finally, the documents are sorted by their ranking scores. The ranking model is trained in advance using learning to rank, and the matching models are trained in advance using semantic matching techniques (or in general learning to match). In this paper, we employ the IR techniques for short text conversation.

\section{Conversation on Social Media}
\label{sec:conversation_example}

\begin{figure}
\centering
\includegraphics[width=.5\textwidth]{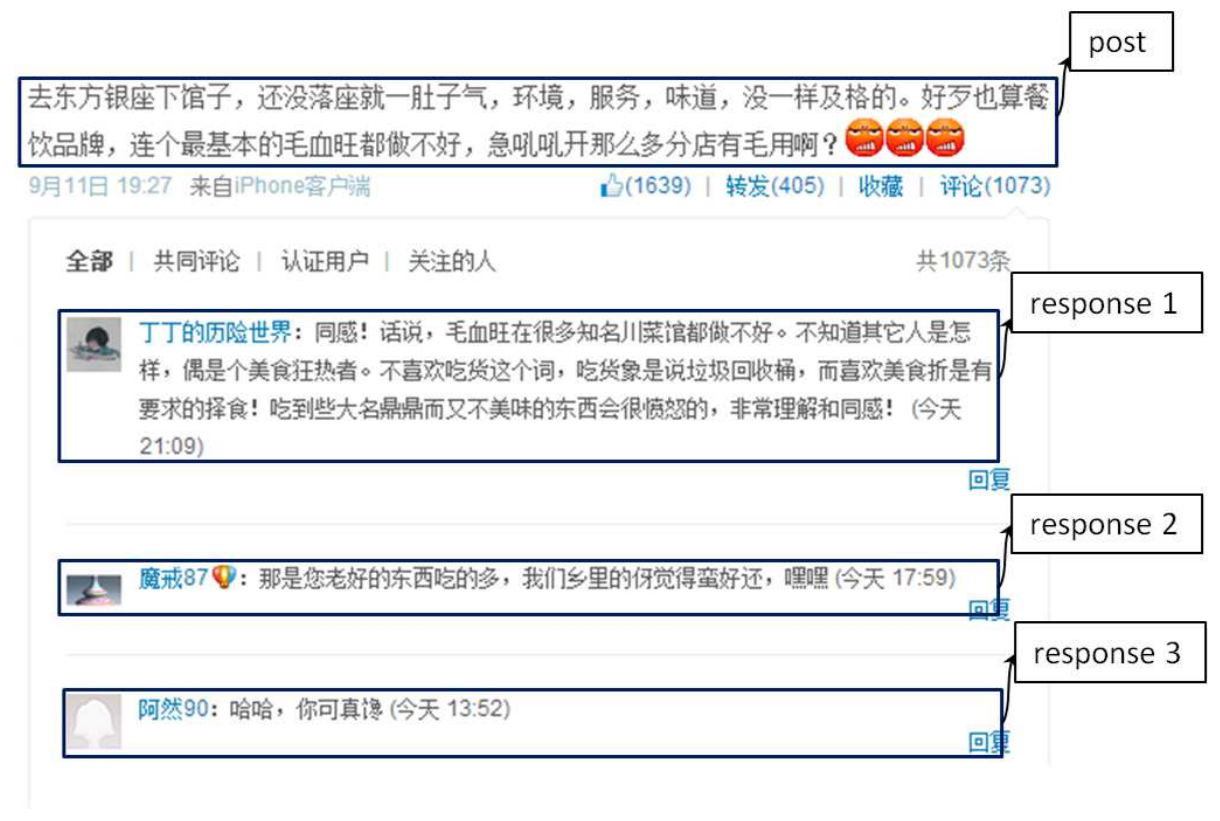}
\caption{An example of post and associated comments at Weibo.}
\label{fig:example_weibo}
\end{figure}

%%\begin{table*}[!h]\small
\begin{table*}\small
  \centering
  \caption{An example of Weibo post and associated comments. The original texts are in Chinese, and we translate them into English. We do the same thing for all the examples in this paper.}\label{tab:conversation_example}

  %\addtolength{\tabcolsep}{-5pt}
  \begin{tabular}{l|p{14cm}}
    \hline

    \textbf{Post} & \\ \hline

    User \textbf{A}: & \tabincell{p{14cm}}{在夏威夷的第一天，端着一大杯葡萄酒，在阳台看日落。 \\ The first day at Hawaii. Watching sunset at the balcony with a big glass of wine in hand.} \\ \hline \hline

    \textbf{Comments} & \\ \hline

    User \textbf{B}: & \tabincell{p{14cm}}{好好享受，别忘了分享照片！ \\ Enjoy it \& don't forget to share your photos!} \\ \hline

    User \textbf{C}: & \tabincell{p{14cm}}{下次带我一起去！ \\ Please take me with you next time!} \\ \hline

    User \textbf{D}: & \tabincell{p{14cm}}{打算在那呆多久？ \\ How long are you going to stay there?} \\ \hline

    User \textbf{E}: & \tabincell{p{14cm}}{你什么时候作报告？ \\ When will be your talk?} \\ \hline

    User \textbf{F}: & \tabincell{p{14cm}}{哈哈，此时我在做着同样的事。你住在哪个酒店？ \\ Haha, I am doing the same thing right now. Which hotel are you staying in?} \\ \hline

    User \textbf{G}: & \tabincell{p{14cm}}{别炫了，老兄！此时我们还在实验室疯狂的编码呢。。 \\ Stop showing-off, buddy. We are still coding crazily right now in the lab.} \\ \hline

    User \textbf{H}: & \tabincell{p{14cm}}{你真幸运！我们去火奴鲁鲁的航班延误了，我被困在了机场。现在在麦当劳嚼着薯条。 \\ Lucky you! Our flight to Honolulu is delayed and I am stuck in the airport. Chewing French fries in MacDonald's right now.} \\ \hline

  \end{tabular}
\end{table*}

Weibo is a microblog service in China, similar to Twitter, on which a user can publish a short message (referred to as \textit{post} in the remainder of the paper) visible to the public or a group of users following her/him. Just like Twitter, Weibo also has the length limit of 140 Chinese characters on each post. Users can attach a short message to a published post, with the same length limit, referred to as \textit{comment} in this paper. Figure \ref{fig:example_weibo} shows an example of post and associated comments (in Chinese).

We argue that the post-comment pairs on Weibo provide a rather valuable resource for studying short text conversation between users. The comments to a post can be of flexible forms and diverse topics, as illustrated in the example in Table \ref{tab:conversation_example}. With a post being a report about the user's current status (travelling to Hawaii), the comments can be a question about the user's future status, a request to the user, a greeting to the user, and so on, but are apparently all appropriate.

In many cases, the post-comment pair is self-contained, which means that one does not need any background knowledge and context information to understand the conversation (Examples of that include the comments of users \textbf{B}, \textbf{D}, \textbf{G} and \textbf{H}). In some cases, one may need extra knowledge and information to understand the conversation. For example, the comment of user \textbf{E} will be fairly elusive if taken out of the context that \textbf{A}'s Hawaii trip is for an international conference and he is going to give a talk there. We argue that the number of self-contained post-comment pairs is vast, and therefore the collected post-comment pairs can serve as a rich resource for exploring rather sophisticated patterns and structures of natural language conversations.

\section{Retrieval-based Short Text Conversation}
\label{sec:framework}

\subsection{Problem Definition}
\textbf{Short text conversation (STC)} is defined as one round of conversation via two short texts, with the former being a message from human and the latter being a response to the message given by the computer.
As the first step, we formalize STC as an information retrieval (IR) problem, i.e., conduct \textbf{retrieval-based STC}. Given a message (query), the system retrieves related responses from the large repository of conversation data and returns the most reasonable response.
With advanced IR technologies and a dataset with previously unthinkable volume, we would expect the conversation system can behave almost like a human in each round of conversation.

Formally, for a given query $q$, we select from the repository of post-comment pairs $(p, r)$ the response $r$ with the highest ranking score.

\begin{equation}
r^{*}=\argmax_{(p,r)}score(q, (p, r))
\end{equation}
where the score is an ensemble of individual matching features.

\begin{equation} \label{equ:rankingsvm_linear}
score(q, (p, r)) = \sum_{i\in\Omega}\omega_{i}\Phi_{i}(q, (p, r))
\end{equation}
where $\Phi_{i}(q, (p, r)$ is the score of the $i$-th matching feature and $\omega_{i}$ is the corresponding feature weight.

\subsection{System Architecture}

\begin{figure}
\centering
\includegraphics[width=.75\textwidth]{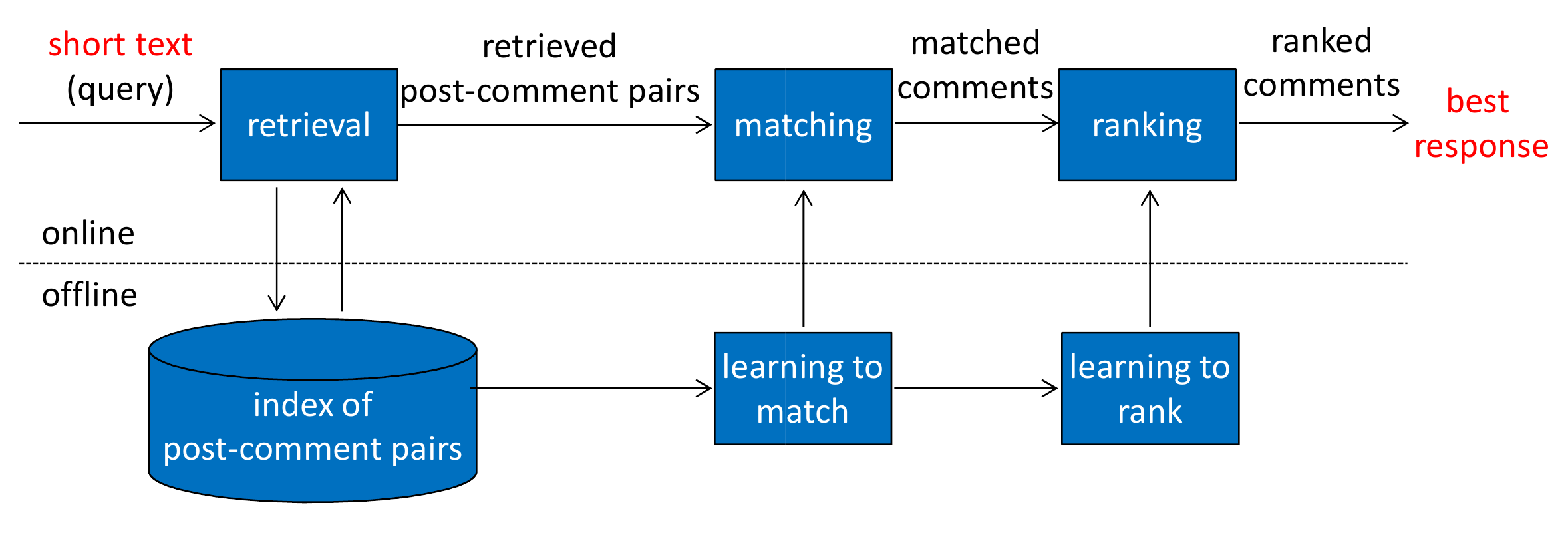}
\caption{System architecture of retrieval-based short text conversation.}
\label{fig:diagram_stc}
\end{figure}

The system performs retrieval-based short text conversation in three-stages, as illustrated in Figure \ref{fig:diagram_stc}:

\begin{itemize}
\item \textbf{Stage I (retrieval)}, the system employs three fast basic linear matching models (see Section \ref{subsec:basic_matching_models}) to retrieve a number of candidate post-comment pairs for the given query $q$, forming a reduced candidate set $C_q^{(reduced)}$.

\item \textbf{Stage II (matching)}, the system utilizes more matching models (see Section \ref{subsec:TransLM} $\sim$ \ref{subsec:simple_matching_features}) to further evaluate all the comments in $C_q^{(reduced)}$, returning a matching feature set $\{\Phi_{i}(q, (p, r), i\in\Omega\}$ for each candidate post-comment pair. The matching models are learned offline with techniques referred to as learning to match (cf., Section \ref{sec:matching_features} for details).

\item \textbf{Stage III (ranking)}, the system uses a linear ranking function defined in Equation (\ref{equ:rankingsvm_linear}) with the matching models as features to further evaluate all the comments (responses) in $C_q^{(reduced)}$, and assigns a ranking score to each candidate comment. Then, the system ranks the candidate comments based on their scores and selects the comment with the highest score to respond. The linear ranking function is learned offline with learning to rank techniques.
\end{itemize}

\subsection{Learning of Ranking Model}

We employ linear RankingSVM \citep{Herbrich:1999:RankingSVM}, a state-of-the-art method of learning to rank, to train the ranking model. We use as training data labeled post-comment pairs, as explained in Section \ref{subsec:labeled_pairs}. From the labeled data, we derive pairwise preference data $(q, (p, r)^+, (p, r)^-)$ such that $score(q, (p, r)^+) > score(q, (p, r)^-)$. Specifically, $(p, r)^+$ are selected from the labeled positive instances with respect to $q$, while $(p, r)^-$ are selected from the labeled negative instances. We have confirmed that the use of labeled negative instances, instead of randomly selected instances, can yield slightly better results.\footnote{This is because the negative instances are collected from the top ranked candidates with several simple retrieval models, and thus they are more indicative of the difference between positive and negative instances.}

\section{Short Text Conversation Dataset}
\label{sec:dataset}
This section introduces the dataset used for retrieval-based STC.\footnote{We have got permission from Weibo to release the dataset for research purpose and it is available at \href{http://data.noahlab.com.hk/conversation/}{http://data.noahlab.com.hk/conversation/}.} There are (1) original post-comment pairs used as the retrieval repository, and (2) labeled post-comment pairs used for training and testing different retrieval models. We give the detail of creation in the following subsections.

\subsection{Original Post-Comment Pairs}

\begin{figure}
\centering
\includegraphics[width=.83\textwidth]{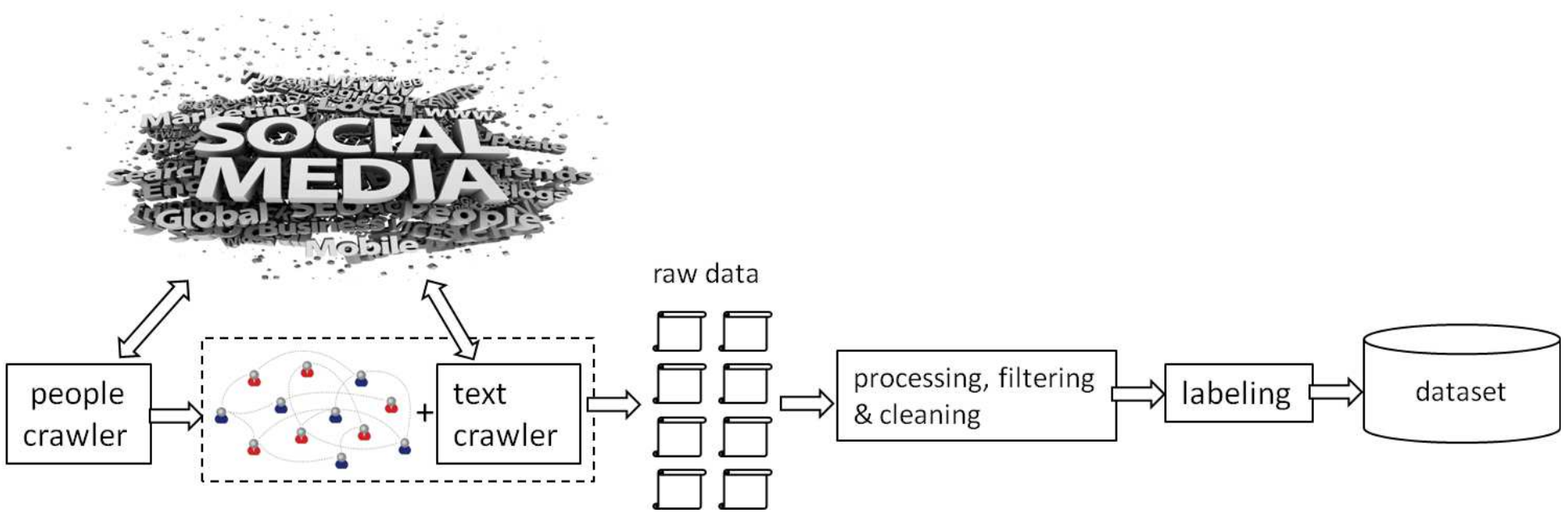}
\caption{Diagram of the process for creating the original and the labeled post-comment pairs.}
\label{fig:diagram_creating_dataset}
\end{figure}

The original post-comment pairs are sampled from Weibo posts published by users in a loosely connected community and the comments they received (may not be from this community).

The creation process of the dataset, as illustrated in Figure \ref{fig:diagram_creating_dataset}, consists of three consecutive steps: (1) crawling the community of users, (2) crawling their posts and the associated comments, and (3) cleaning the data, with more details described below.

\subsubsection{Sampling Strategy}
We take the following sampling strategy for collecting the post-comment pairs to make the topic relatively focused. We first locate 3,200 users from a loosely connected community of Natural Language Processing (NLP) and Machine Learning (ML) in China. The community is mainly composed of professors, researchers, and students. This is done through crawling followees\footnote{When user A follows user B, A is called B's follower, and B is called A's followee.} of ten manually selected seed users who are NLP researchers active on Weibo (with no less than 2 posts per day on average) and popular enough (with no less than 100 followers).

We crawl the posts and the associated comments (not necessarily from the crawled community) for two months (from April 5th, 2013 to June 5th, 2013). The topics are relatively limited due to our choice of the users, with the most saliently ones being:

\begin{itemize}
\item \textbf{Research}: discussion on research ideas, papers, books, tutorials, conferences, and researchers in NLP and ML, etc;
\item \textbf{General Arts and Science}: mathematics, physics, biology, music, painting, etc;
\item \textbf{IT Technology}: mobile phones, IT companies, jobs opportunities, etc;
\item \textbf{Life}: traveling (both touring or conference trips), food, photography, etc.
\end{itemize}

\subsubsection{Processing, Filtering, and Data Cleaning}
\label{subsubsec:preprocess}
On the crawled posts and comments, we first perform a four-step filtering on the posts and comments:

\begin{itemize}
\item We first remove a post-comment pair if the length of the post is less than 10 Chinese characters or the length of the comment is less than 5 Chinese characters. The reason for that is two-fold: (1) if the text is too short, it can barely contain information that can be reliably captured, e.g. the following example

\begin{table}[H]\small
  \centering
\begin{tabular}{|c|l|}
\hline
Post & \tabincell{l}{三个了，还差两个。 \\ Three down, two to go.} \\ \hline
\end{tabular}
\end{table}

and (2) some of the posts or comments are too general to be interesting for other cases, e.g. the comment in the example below

\begin{table}[H]\small
\centering
%\addtolength{\tabcolsep}{-4pt}
\begin{tabular}{|c|l|}
\hline
Post & \tabincell{l}{餐厅不错，强烈推荐。除了排队等待，这里什么都好。 \\ Nice restaurant. I'd strongly recommend it. Everything here is good except the long waiting line.} \\ \hline
Comment & \tabincell{l}{哇！ \\ Wow.} \\ \hline
\end{tabular}
\end{table}

\item In the remained post-comment pairs, we only keep the first 100 comments for each post, since we observe that after the first 100 comments there will be a non-negligible proportion of comments addressing the earlier comments rather than the original post.

\item We then remove the comments that explicitly address other comments from the remained post-comment pairs.

\item The last step is to filter out potential advertisements. We find out long comments that have been posted more than twice on different posts are likely to be advertisements and we scrub them out of the remained post-comment pairs.

\end{itemize}

Then, for the remained posts and comments, we remove punctuation marks and emotions, and use ICTCLAS \citep{Zhang:2003:ICTCLAS} for Chinese word segmentation.

Finally, we get 38,016 Weibo posts and their corresponding 618,104 comments, forming 618,104 \emph{original} post-comment pairs, which are used as the retrieval repository in all the experiments. The statistics of the dataset are summarized in Table \ref{tab:statistics of pairs}.

\begin{table}
\centering
\caption{Statistics of the original and the labeled post-comment pairs.}\label{tab:statistics of pairs}

    \begin{tabular}{|c|l|r|}
    \hline
    \multirow{3}{*}{\tabincell{c}{Original Post-Comment Pairs \\ (Used as retrieval repository)}} & \#posts & 38,016  \\ \cline{2-3}
    \multicolumn{1}{|l|}{}                & \#comments & 618,104 \\ \cline{2-3}
    \multicolumn{1}{|l|}{}                & \#original pairs & 618,104 \\ \hline
    \multirow{3}{*}{\tabincell{c}{Labeled Post-Comment Pairs \\ (Used for training/testing retrieval models)}} & \#posts & 422  \\ \cline{2-3}
    \multicolumn{1}{|l|}{}                & \#comments & 12,402 \\ \cline{2-3}
    \multicolumn{1}{|l|}{}                & \#labeled pairs & 12,402 \\ \hline
    \end{tabular}

\end{table}

\subsection{Labeled Post-Comment Pairs} \label{subsec:labeled_pairs}
This subsection introduces the creation process of the labeled post-comment pairs, which are used for training and testing different retrieval models for STC.

We employ a pooling strategy widely used in information retrieval for getting the instances to label \citep{Voorhees:2002:pooling4ir}.
More specifically, for a given query, we use each of three basic retrieval models to select top 10 comments (see Section \ref{subsec:basic_matching_models} for the description of the basic matching models), and merge them to form a much reduced candidate set with size $\leq$ 30.
Then we assign the comments in the reduced candidate set into ``suitable'' and ``unsuitable'' categories. Basically we consider a comment suitable for a given query if we cannot tell whether it is an original comment. More specifically the suitability of a comment is judged based on the following three criteria\footnote{Note that although our criteria in general favor short and general comments like ``说得很好！(Well said!)'' or ``太好了！(Nice!)'', most of the general comments have already been filtered out due to their lengths (see Section \ref{subsubsec:preprocess}).}:

\begin{itemize}

\item \textbf{Semantic Relevance}: This requires the content of the comment to be semantically relevant to the post. As shown in the example right below, the query is about soccer, and so is comment 1, and hence it is semantically relevant, whereas comment 2 is about food and hence semantically irrelevant.

\begin{table}[H]\small
  \centering
\begin{tabular}{|c|l|}
\hline
Query & \tabincell{l}{英格兰 禁区 里 老 是 八 个 人 … … \\ There are always 8 English players in their own penalty area. Unbelievable!} \\ \hline
Comment 1 & \tabincell{l}{哈哈哈 仍然 是 0：0 。 还 没 看到 进球 。 \\ Haha, it is still 0:0, no goal so far.} \\ \hline
Comment 2 & \tabincell{l}{英格兰 的 食物 真 可怕! \\ The food in England is horrible.} \\ \hline
\end{tabular}
\end{table}

Another important aspect of semantic relevance is entity association. This requires that the entities in the comment to be strongly associated with those in the query. In other words, if the query is about entity A, while the comment is about entity B, they are very likely to be mismatched. As shown in the following example, where the query is about Paris, while comment 2 talks about London:

\begin{table}[H]\small
  \centering
\begin{tabular}{|c|l|}
\hline
Query & \tabincell{l}{巴黎 的 天空 ！ 暖暖 的 太阳 + 阵阵 的 微风 ， 真 让 人 不 舍得 离开 。 \\ The sky in Paris! Warm sun + bursts of breeze. Really hate to leave.} \\ \hline
Comment 1 & \tabincell{l}{好好 享受 你 在 巴黎 的 时间 吧。 \\ Enjoy your time in Paris.} \\ \hline
Comment 2 & \tabincell{l}{我 希望 我 现在 在 伦敦 。 \\ Man, I wish I am in London right now.} \\ \hline
\end{tabular}
\end{table}

This is however not absolute, since a comment containing a different entity could still be sound, as demonstrated by the following two comments to the query above:

\begin{table}[H]\small
  \centering
\begin{tabular}{|c|l|}
\hline
Comment 1 & \tabincell{l}{好好 享受 你 在 法国 的 时间 吧。 \\ Enjoy your time in France.} \\ \hline
Comment 2 & \tabincell{l}{伦敦 的 秋天 也 很 漂亮 。 \\ The fall of London is nice too.} \\ \hline
\end{tabular}
\end{table}

\item \textbf{Logic Consistency}: This requires the content of the comment to be logically consistent with the query. For example, in the table right below, the query states that the Huawei mobile phone ``Honor'' is already in the market of mainland China. Comment 1 talks about a personal preference over the same phone model (hence logically consistent), whereas comment 2 asks a question to which the answer is already clear from the query (hence logically inconsistent).

\begin{table}[H]\small
  \centering
\begin{tabular}{|c|l|}
\hline
Query & \tabincell{l}{华为 荣耀 手机 在 中国 大陆 大 卖 ！ \\ HUAWEI's mobile phone, Honor, sells well in mainland China.} \\ \hline
Comment 1 & \tabincell{l}{华为 荣耀 手机 是 我 最 喜欢 的 一 款 手机。 \\ HUAWEI Honor is my favorite phone} \\ \hline
Comment 2 & \tabincell{l}{华为 荣耀 手机 什么 时候 会 在 中国 大陆 上市？ \\ When will HUAWEI Honor get to the market in mainland China?} \\ \hline
\end{tabular}
\end{table}

\item \textbf{Speech Act Agreement}: Another important factor in determining the suitability of a comment is speech act. For example, when a question is posed in the Weibo post, a certain act (e.g., answering or forwarding it) is expected. In the example below, the query asks a special question about location. Comment 1 forwards and comments 2 answers the question, whereas comment 3 is an imperative sentence and therefore does not correspond well in speech act.

\begin{table}[H]\small
  \centering
\begin{tabular}{|c|l|}
\hline
Query & \tabincell{l}{有人 知道 后年 的 KDD 会 在 哪里 举行 呢 ？ \\ Any one knows where KDD will be held the year after next?} \\ \hline
Comment 1 & \tabincell{l}{同 问 。 希望 在 欧洲 \\ co-ask. Hopefully Europe} \\ \hline
Comment 2 & \tabincell{l}{听说 在 纽约 \\ New York, as I heard} \\ \hline
Comment 3 & \tabincell{l}{请 帮忙 扩散 这条 关于 KDD 的 信息 。 \\ Please help me distribute the information of KDD.} \\ \hline
\end{tabular}
\end{table}

\end{itemize}

Finally, we manually label 422 queries and their associated comments, about 30 comments for each post. Note that (1) the labeling is only on a small subset of the 38,016 posts, and (2) for each selected (query) post, the labeled comments are \emph{not originally} given to it. The statistics of this part of dataset are also summarized in Table \ref{tab:statistics of pairs}.

\section{Matching Features}
\label{sec:matching_features}

In this section, we introduce matching features (calculated from matching models) used in our retrieval-based STC. Three basic linear matching models are first introduced as the baselines. Then, a translation-based language model (TransLM) is introduced for alleviating the lexical gap problem.  Next, a deep matching model (DeepMatch) is described, which matches query and response (comment) with a deep architecture. After that, a topic-word model (TopicWord) is explained for addressing the topic alignment of query and response (comment). Finally, other simple matching features used in our retrieval-based STC are described.

\subsection{Basic Linear Matching Models}
\label{subsec:basic_matching_models}

We use the following three basic linear matching models for fast retrieval in \textbf{Stage I}. Moreover, these matching models are also used in \textbf{Stage II} to generate three matching features for each post-comment pair.

\subsubsection{Query-Response Similarity} \label{subsubsec:Q2R_sim}
Here we use a simple vector space model for measuring the similarity between a query $q$ and a candidate response $r$

\begin{equation} \label{equ:sim_Q2R}
sim_{Q2R}(\mathbf{q}, \mathbf{r})=\dfrac{\mathbf{q}^{\mathsf{T}}\mathbf{r}}{\Vert\mathbf{q}\Vert\Vert\mathbf{r}\Vert}
\end{equation}
where $\mathbf{q}$ and $\mathbf{r}$ are respectively the TF-IDF vectors of $q$ and $r$.

Although it is not necessarily true that a good response has many common words as the query, but this measurement is often helpful in finding relevant responses. For example, when the query and the candidate response both have ``National Palace Museum in Taipei'', it is a strong signal that they are about similar topics. Unlike other semantic matching features, this simple similarity requires no learning and works on infrequent words. Our empirical results show that it can often capture the query-response relation which semantic matching features cannot.

\subsubsection{Query-Post Similarity} \label{subsubsec:Q2P_sim}
The basic idea here is to find posts (messages) similar to the query  $q$ and use their comments (responses) as the candidates. Again we use the vector space model for measuring the query-post similarity

\begin{equation} \label{equ:sim_Q2P}
sim_{Q2P}(\mathbf{q}, \mathbf{p})=\dfrac{\mathbf{q}^{\mathsf{T}}\mathbf{p}}{\Vert\mathbf{q}\Vert\Vert\mathbf{p}\Vert}
\end{equation}
where $\mathbf{q}$ and $\mathbf{p}$ are respectively the TF-IDF vectors of $q$ and $p$.

The assumption here is that if a post $p$ is similar to the query $q$, its associated comments (responses) might be appropriate for $q$. It however often fails, especially when a response to $p$ addresses parts of $p$ not contained by $q$, which fortunately can be alleviated when combined with other measures.

\subsubsection{Query-Response Matching in Latent Space} \label{subsubsec:semantic_matching}
This particular matching function relies on mapping of posts and responses in the original vector spaces to a low-dimensional latent space, learned from data. The matching score between a query $q$ and a candidate response $r$ can be measured as the inner product between their images in the latent space.

\begin{equation}
LatentMatch(\mathbf{q}, \mathbf{r})=\mathbf{q}^{\mathsf{T}}L_{\mathbf{q}}L_{\mathbf{r}}^{\mathsf{T}}\mathbf{r}
\end{equation}

This is to capture the semantic matching between a post and a response, which may not be well captured by a word-to-word matching. We find the mapping functions $L_{\mathbf{q}}$ and $L_{\mathbf{r}}$ through a large number of query-response pairs and a large margin variant of the method in \citet{Wu:2013:bilinear}.

\begin{equation}
\begin{aligned}
\argmin_{L_{\mathbf{q}},L_{\mathbf{r}}}~& \sum_{i}\max(1-\sum_{i}\mathbf{q}_{i}^{\mathsf{T}}L_{\mathbf{q}}L_{\mathbf{r}}^{\mathsf{T}}\mathbf{r}_{i}, 0) \\
\text{s.t.}~& \Vert L_{n,\mathbf{q}}\Vert_{1}\leq \mu_{1}, n=1,2,...,N_{\mathbf{q}}\\
& \Vert L_{m,\mathbf{r}}\Vert_{1}\leq \mu_{1}, m=1,2,...,N_{\mathbf{r}}\\
& \Vert L_{n,\mathbf{q}}\Vert_{2}= \mu_{2}, n=1,2,...,N_{\mathbf{q}}\\
& \Vert L_{m,\mathbf{r}}\Vert_{2}= \mu_{2}, m=1,2,...,N_{\mathbf{r}}
\end{aligned}
\end{equation}
where $i$ indices the original post-comment pairs.

Our experiments (see Table \ref{tab:Example_effect_LatentMatch}) indicate that this simple linear model can learn meaningful semantic matching patterns, due to its effective use of massive data.
For example, the image of the word ``Italy'' in the post in the latent space matches well the words ``Sicily'', ``Mediterranean sea'' and ``travel''. Once the mapping $L_{\mathbf{q}}$ and $L_{\mathbf{r}}$ are learned, the semantic matching score $\mathbf{q}^{\mathsf{T}}L_{\mathbf{q}}L_{\mathbf{r}}^{\mathsf{T}}\mathbf{r}$ will be utilized as a feature for modeling the overall suitability of $r$ as a response to $q$.

\subsection{Translation-based Language Model}\label{subsec:TransLM}

\subsubsection{Motivation}
\comment{motivation}
With the three basic matching models and some simple features (see Section \ref{subsec:simple_matching_features}), the model for retrieval-based STC can achieve fairly good performance (see Section \ref{subsec:results_basic}). However, there are also some cases in which the model fails. One of the serious problems is the \textit{lexical gap}, which is the major challenge for most information retrieval tasks, between the query and the candidate post-comment pairs. Table \ref{tab:Example of lexical gap problem} shows a real example of the lexical gap problem. Two candidate responses are suitable to the query, while their ranking is very low in the top 30 candidates. The main reason is that there is no word overlap between the candidate responses and the query, although there is one common word ``晚安(Good Night)'' between the original posts and the query.

\comment{example}
%%\begin{table*}\small
\begin{table*}[!htb]\small
  \centering
  \caption{An example of the lexical gap problem between the query and the candidate post-comment pairs. The ``Labels'' column indicates whether the candidate responses are suitable to the query. The ``Candidate Responses'' and ``Original Posts'' columns list the candidate responses and their original posts, respectively. The ``Rank'' column shows the ranking of the 2 responses in the top 30 candidates with the basic matching models.}\label{tab:Example of lexical gap problem}

    %\addtolength{\tabcolsep}{-4pt}
    \begin{tabular}{|l|p{8cm}|p{4cm}|c|}
    \hline

    \textbf{Query} & \multicolumn{3}{p{11cm}|}{\tabincell{p{14.1cm}}{小舟 划向 夕阳 （ 摄于 西西里岛 附近 ） 。 晚安 ！ \\ A boat rowed off into the sunset (taken in the vicinity of Sicily). Good Night!}} \\
    \hline \hline

    \textbf{Labels} & \textbf{Candidate Responses} & \textbf{Original Posts} & \textbf{Rank} \\ \hline

    Suitable & \tabincell{p{8cm}}{日落 夜 不 黑 的 欧洲 夜晚 \\ Sunset night, not black night in Europe} & \tabincell{p{4cm}}{苏黎世 日落 。 大家 晚安 ！ \\ Zurich sunset. Good night!} & 22 \\ \hline

    Suitable & \tabincell{p{8cm}}{种么 美 ， 想起 《 爱 在 黄昏 日落 时 》 的 场景 \\ What a beauty! I think of the ``Love in the Sunset'' scene} & \tabincell{p{4cm}}{巴黎 日落 。 大家 晚安 ！ \\ Paris sunset. Good night!} & 25 \\ \hline

    \end{tabular}

\end{table*}

\subsubsection{Model Description}
To alleviate the lexical gap problem, we employ the state-of-the-art translation-based language model (TransLM) \citep{Xue:2008:TransLM} with a small modification for retrieval-based STC. Given a query  $q$ and a candidate post-comment pair $(p,r)$, the ranking  function based on TransLM is written as

\begin{equation}
P_{TransLM}(q|(p,r))=\prod_{w\in q}P_{TransLM}(w|(p,r))
\end{equation}

\begin{equation} \label{equ:alpha}
P_{TransLM}(w|(p,r))=(1-\alpha)P_{mx}(w|(p,r))+\alpha P_{ml}(w|C)
\end{equation}

\begin{equation} \label{equ:beta_gamma}
\begin{split}
P_{mx}(w|(p,r))=&(1-\beta)\left[(1-\gamma)P_{ml}(w|p)+\gamma \sum_{t\in p}T(w|t)P_{ml}(t|p)\right]\\
&+\beta\left[(1-\gamma)P_{ml}(w|r)+\gamma\sum_{t\in r}T(w|t)P_{ml}(t|r)\right]
\end{split}
\end{equation}
where $P_{ml}(w|p)$, $P_{ml}(w|r)$, and $P_{ml}(w|C)$ are the unigram language models (LM), which are estimated with maximum likelihood, for the post part $p$, the response part $r$ and the whole collection $C$, respectively.
$T(w|t)$ is the probability of translating a word $t$ in $p$ or $r$ into a word $w$ in $q$.
$\sum_{t\in p}T(w|t)P_{ml}(t|p)$ and $\sum_{t\in r}T(w|t)P_{ml}(t|r)$ are the translation models (Trans) for the post part and response part, respectively.
\comment{Why TransLM can alleviate the lexical gap problem?}
To bridge the lexical gap between queries and candidate post-comment pairs, the two translation models allow a post and its response translate any one of their words $t$ to a different but semantically related query word $w$ with a non-zero probability.
$\alpha$ is the Jelinek-Mercer smoothing factor \citep{Zhai:2001:Smoothing}. $\beta$ is the interpolation parameter between the post model and the response model. $\gamma$ is the parameter to balance between the unigram language model and the translation model for alleviating the lexical gap problem and the self-translation problem \citep{Xue:2008:TransLM}.

\comment{
If we set $\beta=0$, only the post part is used for computing the ranking score; if we set $\beta=1$, only the response part is used for computing the ranking score; otherwise, we combine the ranking scores computed from both parts.

If we set $\gamma=0$, only the unigram language model is used for computing the ranking score; if we set $\gamma=1$, only the translation model is used for computing the ranking score; otherwise, we combine the ranking score computed from both models.
}

The main difference between our TransLM for retrieval-based STC and that for question retrieval in Community Question Answering (CQA) \citep{Xue:2008:TransLM} is that we add an additional part $\sum_{t\in r}T(w|t)P_{ml}(t|r)$ in the response part while \citet{Xue:2008:TransLM} does not do that in the answer part. The reasons are two-fold: (1) we focus on finding suitable responses given a query while \citet{Xue:2008:TransLM} focus on finding similar questions given a query question, and (2) the responses tend to be shorter than the posts in STC while the answers tent to be longer than the questions in CQA.

\subsubsection{Learning Translation Probabilities} \label{subsubsec:learning_trans}
The performance of the translation-based language model relies on the quality of the word-to-word translation probabilities. We follow the method of \citet{Xue:2008:TransLM} to learn the word translation probabilities. In our experiments, original post-comment pairs are used for training, and the GIZA++\footnote{https://code.google.com/p/giza-pp/} \citep{Och:2003:GIZA++} toolkit is used to learn the translation model. There are two different settings of source and target for constructing the parallel corpus: (1) set the posts as the source and the responses as the target, i.e., collection ${(p,r)_1, ..., (p,r)_n}$, and (2) set the responses as the source and the posts as the target, i.e., collection ${(r,p)_1, ..., (r,p)_n}$. Following \citet{Xue:2008:TransLM}, a pooling strategy is adopted to combine the above two collections of pairs to get a pooled collection ${(p,r)_1, ..., (p,r)_n, (r,p)_1, ..., (r,p)_n}$. Moreover, we also filter low-frequency words, whose frequency is less than 10.

\subsection{Deep Matching Model}\label{subsec:DeepMatch}

\subsubsection{Motivation}

The matching models above are linear models. These models, although proven to be effective, are insufficient for capturing the rich structures in matching complicated objects like texts. We propose employing a new deep neural network model, referred to it as deep matching model (DeepMatch) in the paper, to model the complicated matching relations between query and candidate response in retrieval-based STC.\footnote{The detail can be found in \citet{Lu:2013:DeepMatch}.} This new architecture is mainly based on the following two intuitions:

\begin{itemize}
\item \textbf{Localness}: There are salient local structures in the semantic space of parallel texts to be matched, which can be roughly captured by the co-occurrence pattern of words. This localness however should not prevent two ``distant'' components from correlating with each other on a higher level, hence call for the hierarchical characteristic of the model;

\item \textbf{Hierarchy}: The decision making for matching has different levels of abstraction. The local decisions, capturing the interactions between semantically related words, will be combined later layer-by-layer to form the final and global decision on matching.
\end{itemize}

\subsubsection{Model Description}

\begin{figure}
\centering
\includegraphics[width=.65\textwidth]{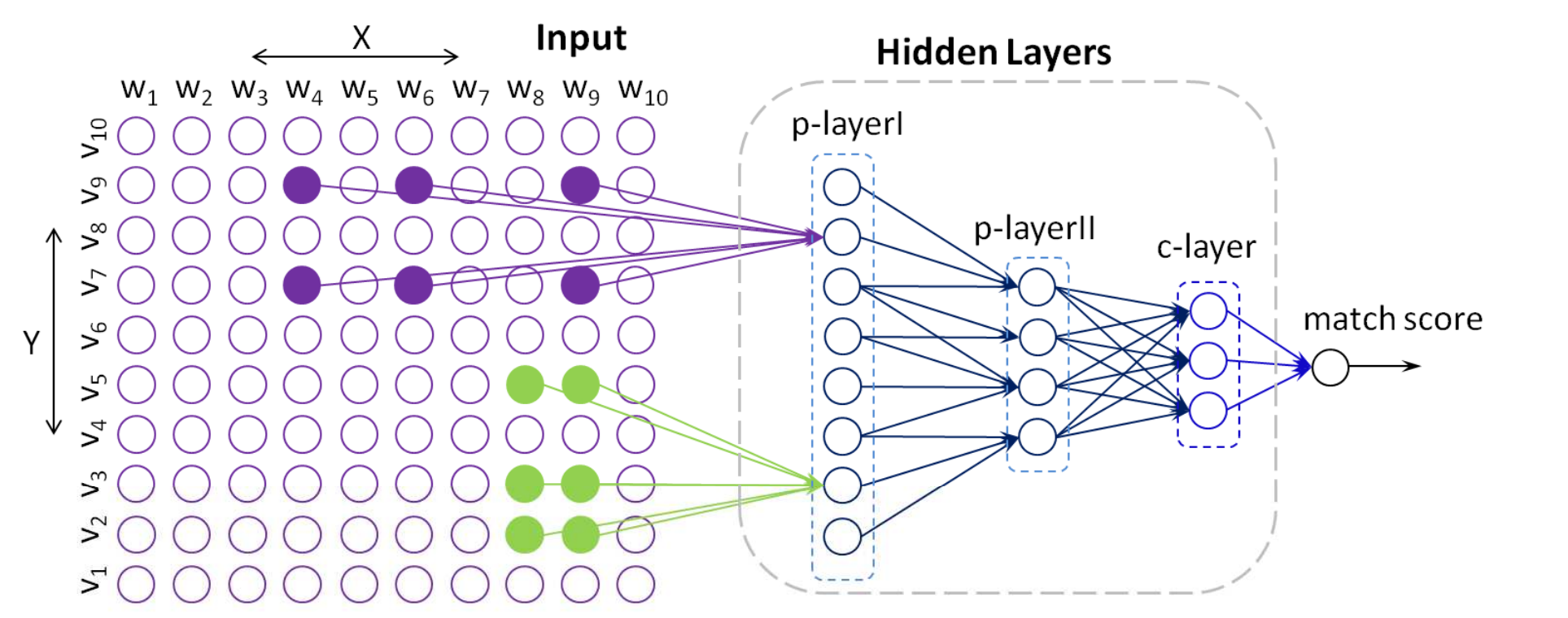}
\caption{An illustration of the architecture of deep matching model.}
\label{fig:DeepMatch}
\end{figure}

The deep matching model (DeepMatch) consists of two parts: (1) many bilinear local matching models, and (2) a deep neural network to further combine the local matching models to generate the final matching score. Each local matching model (indexed $k$) is in charge of a small subset of words in both short texts ($\mathbf{x}$ and $\mathbf{y}$) and a pair of projection matrices $(L^{(k)}_{x},\; L^{(k)}_{y})$. The score from the $k^{th}$ matching model is given as follows
\begin{equation}
a^{(k)}(\mathbf{x},\mathbf{y}) =  f^{(k)}\left((\mathbf{x}^{(k)})^\top L^{(k)}_{x} (L^{(k)}_{y})^\top \mathbf{y}^{(k)} + b^{(k)} \right), \;\; k = 1,\cdots,K
%\vspace{-5pt}
%= \sigma(a_p(\mathbf{x},\mathbf{y}))
\end{equation}
where $f^{(k)}(\cdot)$ is a function with sigmoid shape for mapping the matching score to $(0, 1)$. Those local matching decisions are then used as input to feed into a multi-layer perceptron to calculate the final matching score. The overall architecture of DeepMatch is given in Figure \ref{fig:DeepMatch}.

\subsubsection{Model Training}

The training of DeepMatch is divided into two phases: (1) bilingual topic modeling for finding potentially matched subsets (topics) of word-pairs and building the model architecture, and (2) training of the parameters of the local topic models and deep neural network. In this paper, we train the deep matching model with a ranking-based objective. More specially, we employ a large margin objective defined on preference pairs in ranking. Suppose that we are given the following triples $(\mathbf{x}, \mathbf{y}^+, \mathbf{y}^-)$ from the oracle, with $\mathbf{x}$ ($\in \mathcal{X}$) matched with $\mathbf{y}^+$ better than with $\mathbf{y}^-$ (both $\in \mathcal{Y}$).  The ranking-based objective is defined as follows\vspace{-2pt}

\begin{equation}
\mathcal{L}(\mathcal{W}, \mathcal{D}_{trn}) = \sum_{(\mathbf{x}_i, \mathbf{y}_{i}^+, \mathbf{y}_{i}^-) \in \mathcal{D}_{trn}} e_{\mathcal{W}}(\mathbf{x}_i, \mathbf{y}_{i}^+, \mathbf{y}_{i}^-) + R(\mathcal{W}), \vspace{-5pt}
\end{equation}
where $R(\mathcal{W})$ is the regularization term and $e_{\mathcal{W}}(\mathbf{x}_i, \mathbf{y}_{i}^+, \mathbf{y}_{i}^-)$ is the error for triple $(\mathbf{x}_i, \mathbf{y}_{i}^+, \mathbf{y}_{i}^-)$, given by the following large margin form:
\[
e_i=e_{\mathcal{W}}(\mathbf{x}_i, \mathbf{y}_{i}^+, \mathbf{y}_{i}^-) =
%\begin{cases}
\max(0, m+\textsf{s}(\mathbf{x}_i,\mathbf{y}_i^-)-\textsf{s}(\mathbf{x}_i,\mathbf{y}_i^+)),
%\end{cases}
\]
with parameter $0<m$ controlling the margin. In the experiments we use $m=2$, but we find that the results are rather insensitive to the values of $m$ varying in a fairly large range.

\subsection{Topic-Word Model}
\label{subsec:TopicWord}

\subsubsection{Motivation}

\comment{motivation}
The matching models above are mainly for representing semantic relevance between query and response. They do not capture the matching relations between the main topics of the query and the response. Table \ref{tab:Example of topic words problem} shows a real example of the problem. The top 2 candidate responses are unsuitable to the query, while the first suitable response is ranked at 9. The main reason is that the word ``菜鸟(rookie)'' has higher term frequencies in the query and the unsuitable responses, and thus dominates their matching scores. However, the main topics of the query is not ``菜鸟(rookie)'' but ``代码 控制 工具 (code control tool) SVN GIT'', and thus the top 2 candidate responses are not suitable. One possible solution is to identify the topic words of the query and the candidate responses and give higher weights to the matching of the topic words. This may alleviate the topic mismatch problem and improve the performance of STC. In Table \ref{tab:Example of topic words problem}, the topic words of the query and the candidate responses are shown in bold. After identifying the topic words, the first two responses clearly become unsuitable to the query, while the third response having the same topic becomes suitable.

One key question is how to differentiate topic words from the other words in a short text. We propose a method for the task in the following subsections.

\comment{example}
%%\begin{table*}[!htb]\small
\begin{table*}\small
  \centering
  \caption{An example to show that matching of topic words is necessary for retrieval-based STC. The ``Labels'' column indicates whether the candidate responses are suitable to the query. The ``Rank'' column shows the ranking of the 3 responses in the top 30 candidates with the basic matching models. The words in bold are the topic words of the post and responses.}\label{tab:Example of topic words problem}
    %\addtolength{\tabcolsep}{-4pt}
    \begin{tabular}{|l|p{12.8cm}|c|} \hline

    \textbf{Query} & \multicolumn{2}{p{12.8cm}|}{\tabincell{p{14.1cm}}{选用 \textbf{代码 控制 工具} 的 时候 ， 还是 用 \textbf{SVN} ， 不 要 用 \textbf{GIT} ， 虽然 \textbf{GIT} 比 \textbf{SVN} 强大 很多 。 因为 如果 团队 里 有 一 个 菜鸟 ， 他 各种各样 关于 \textbf{GIT} 的 问题 会 烦 死 你 。 \textbf{SVN} 的 优势 就 是 极其 简单 ， 菜鸟 也 可以 很 快 掌握 \\ When choosing the \textbf{code control tool}, use \textbf{SVN} instead of \textbf{GIT}, although \textbf{GIT} is more powerful than \textbf{SVN}. The reason is that if a team has a rookie, he will bother you with all sorts of questions about \textbf{GIT}. The advantage of \textbf{SVN} is it is so simple that a rookie can grasp it quickly enough.}} \\ \hline \hline

    \textbf{Labels} & \textbf{Candidate Responses} & \textbf{Rank} \\ \hline

    Unsuitable & \tabincell{p{12.8cm}}{我 是 \textbf{保密员} 、 \textbf{菜鸟} 与 \textbf{90后} \textbf{努力} 为了 不 成为 老 \textbf{菜鸟} \\ I am a \textbf{confidential staff}, \textbf{rookie} and \textbf{the generation after 90s}, \textbf{making efforts} in order not to become an old \textbf{rookie}. } & 1 \\ \hline

    Unsuitable & \tabincell{p{12.8cm}}{我 ， 还 有 很多 用户 。 都 是 \textbf{菜鸟} 。 \\ I and also a lot of users are \textbf{rookies}.} & 2 \\ \hline

    Suitable & \tabincell{p{12.8cm}}{一 个 \textbf{版本 控制} 的 \textbf{工具} \\ A \textbf{version control tool}} & 9 \\ \hline

    \end{tabular}

\end{table*}

\subsubsection{Learning Topic Words}
\comment{definition of topic words in a short text}

A short text, such as a post or a comment in Weibo, usually centers around a specific theme, which is usually captured by a number of words in the text. We refer to the words as \textit{topic words}. Examples are show in Table \ref{tab:Example of topic words problem}.

\comment{learning topic words}
In this paper, we employ a probabilistic approach based on logistic regression to compute the probability of a word being the topic word of a short text. In the logistic regression model, we specifically compute $P(topic|w)$ as follows, where $w$ denotes a word and $topic$ denotes a binary variable representing whether or not $w$ is a topic word.

\begin{equation} \label{equ:logistic_regression}
\log\left(\dfrac{P(topic|w)}{1-P(topic|w)}\right)=\vec{\omega}\cdot\vec{x}+c
\end{equation}

\begin{equation} \label{equ:p_topic_give_w}
P(topic|w)=\dfrac{e^{\vec{\omega}\cdot\vec{x}+c}}{1+e^{\vec{\omega}\cdot\vec{x}+c}}
\end{equation}
where $\vec{x}$ is a short text, such as a post or a comment in Weibo, represented as a vector of features, $\vec{\omega}$ is a vector of weights associated with the features, and $c$ is a constant.

In the logistic regression model we make use of the features, as listed in Table \ref{tab:features}. The first two features (TF, IDF) are from the traditional term weighting schemes in information retrieval. The third feature (SF) is based on the observation that words appearing in multiple sentences in the short text are more likely to be topic words. The next two features (First, Last) represent the positions of word in the short text. Our observation is that topic words are most likely to appear in the first or the last sentence. The next three features (NE, NE\_First, NE\_Last) are named entity related features. The last feature (POS) is the part of speech of a word, which is based on the fact that topic words are usually denoted by nouns and verbs.

\begin{table*}
  \centering
  \caption{Features for predicating the probability of word $w$ being a topic word in a short text.}\label{tab:features}

  \begin{tabular}{|c|l|l|}
    \hline

    \textbf{\#} & \textbf{Features} & \textbf{Description} \\ \hline\hline

    1 & TF & Term frequency of $w$ in the short text \\ \hline
    2 & IDF & Inverse document frequency of $w$ in the whole collection \\ \hline
    3 & SF & Number of sentences in the short text that contain $w$ \\ \hline
    4 & First & Whether $w$ exists in the first sentence \\ \hline
    5 & Last & Whether $w$ exists in the last sentence \\ \hline
    6 & NE & Whether $w$ is a named entity (NE) \\ \hline
    7 & NE\_First & Whether $w$ is NE in the first sentence \\ \hline
    8 & NE\_Last & Whether $w$ is NE in the last sentence \\ \hline
    9 & POS & Part of speech of $w$ \\ \hline
  \end{tabular}

\end{table*}

\comment{Labeled Topic Words}
To create the training data, we randomly select 200 short texts (including posts and comments) to label their topic words. We make labeling judgments on all words in a short text (except some stop words), as our model is applied to words. Each word is assigned to one of the two classes, ``positive'' class and ``negative'' class, depending on whether or not it is a topic word. The judgments are made based on the principle that the topic words of a short text should be indicative to the main theme of the short text.
\comment{statistics of the labeled data}
Finally, we obtain 2,008 words for the 200 short texts. Table \ref{tab:statistics_of_labeled_topic_words} summarizes the statistics of the dataset and Table \ref{tab:topic_word_distribution} lists the topic word distributions over word positions, parts of speech and named entities in the dataset. As shown in Table \ref{tab:topic_word_distribution}, topic words are typically nouns in the first sentence of a short text.

%%\begin{table*}[!htb]\small
\begin{table*}\small
\centering
\caption{Statistics of topic words.} \label{tab:statistics_of_labeled_topic_words}
\begin{tabular}{|c|c|c|}
\hline

\#topic words & \#non-topic words & \#total words \\ \hline
847 & 1,161 & 2,008 \\ \hline

\end{tabular}
\end{table*}

%%\begin{table*}[!htb]\small
\begin{table*}\small
\centering
\caption{Topic word distributions over word positions, parts of speech and named entities.} \label{tab:topic_word_distribution}
\begin{tabular}{|c|c|r|}
\hline

\multirow{3}{*}{Topic word distribution over word positions} & In the first sentence & 83.33\% \\ \cline{2-3}
\multicolumn{1}{|c|}{}  & In the last sentence & 15.12\% \\ \cline{2-3}
\multicolumn{1}{|c|}{}  & In the other sentences & 1.54\% \\ \hline \hline

\multirow{4}{*}{Topic word distribution over parts of speech (POS)} & Noun & 60.33\% \\ \cline{2-3}
\multicolumn{1}{|c|}{}  & Verb & 26.33\% \\ \cline{2-3}
\multicolumn{1}{|c|}{}  & Adjective & 8.61\% \\ \cline{2-3}
\multicolumn{1}{|c|}{}  & Others & 4.72\% \\ \hline \hline

\multirow{2}{*}{Topic word distribution over named entities (NE)} & Is NE & 12.04\% \\ \cline{2-3}
\multicolumn{1}{|c|}{}  & Not NE & 87.96\% \\ \hline

\end{tabular}
\end{table*}

We use ICTCLAS \citep{Zhang:2003:ICTCLAS} to obtain the part of speech and named entity information and LIBLINEAR \citep{Fan:2008:LIBLINEAR} to build the logistic regression model and predict the probability of a word being a topic word. The accuracy of our model is 81.57\%.

\comment{use the learned topic words for retrieval}
\subsubsection{Model Description}
With the trained model, we can assign a probability value to each word in the short text, which indicates the probability of the word being a topic word. The question is how to use this information in retrieval-based STC. In this work, we simply take the probability values as term weights and use them to calculate the similarities between a query $q$ and a candidate response $r$ or its original post $p$ with a vector space model, which are similar to the query-response similarity and query-post similarity introduced in Section \ref{subsec:basic_matching_models}.

\begin{equation} \label{equ:sim_Q2R_TopicWord}
sim_{Q2R\_TopicWord}(\mathbf{q}, \mathbf{r})=\dfrac{\mathbf{q}^{\mathsf{T}}\mathbf{r}}{\Vert\mathbf{q}\Vert\Vert\mathbf{r}\Vert}
\end{equation}

\begin{equation} \label{equ:sim_Q2P_TopicWord}
sim_{Q2P\_TopicWord}(\mathbf{q}, \mathbf{p})=\dfrac{\mathbf{q}^{\mathsf{T}}\mathbf{p}}{\Vert\mathbf{q}\Vert\Vert\mathbf{p}\Vert}
\end{equation}
where $\mathbf{q}$, $\mathbf{p}$ and $\mathbf{r}$ are respectively the vectors of $q$, $p$ and $r$ with topic word probability $P(topic|w)$ as weight.

\comment{
Moreover, we further interpolate these two matching scores to combine their strength.

\begin{equation} \label{equ:sim_Q2R+Q2P_TopicWord}
sim_{TopicWord}(q, (p,r)) = (1-\lambda)sim_{Q2R\_TopicWord}(\mathbf{q}, \mathbf{r}) + \lambda sim_{Q2P\_TopicWord}(\mathbf{q}, \mathbf{p})
\end{equation}
where $\lambda$ is the interpolation parameter.
}

\subsection{Other Simple Matching Features} \label{subsec:simple_matching_features}
We also use some other simple matching features as follows:

\begin{itemize}
\item Longest Common String (LCS): this feature measures the length of the longest common string between the query $q$ and the candidate response $r$, which is useful for capturing the quotes common in microblog comments and is also fairly robust to errors in Chinese word segmentation.

\begin{equation}
LCS_{Q2R}(q, r)=|LCS(q,r)|
\end{equation}

\item Co-occurrence features: these features represent the size, the rate, the sum, and the average of IDF values of the co-occurring words between the query $q$ and the candidate response $r$ or its original post $p$.

\begin{equation}
cooccur\_size(x, y) = |cooccur(x,y)|
\end{equation}

\begin{equation}
cooccur\_rate(x, y) = \dfrac{|cooccur(x,y)|}{|y|}
\end{equation}

\begin{equation}
cooccur\_sumIDF(x, y) = \sum_{w\in cooccur(x,y)}IDF(w)
\end{equation}

\begin{equation}
cooccur\_averageIDF(x, y) = \dfrac{\sum_{w\in cooccur(x,y)}IDF(w)}{|cooccur(x,y)|}
\end{equation}
where $x$ stands for the query $q$ and $y$ stands for the candidate response $r$ or its original post $p$, $cooccur(x,y)$ is the set of common words between $x$ and $y$.

\end{itemize}

\section{Experiments}
\label{sec:experiment}
We conduct experiments and report results on retrieval-based STC with the dataset we have created.

\subsection{Experimental Setup}

\subsubsection{Evaluation Metrics}
We evaluate the performance of different retrieval models for STC based on the following two metrics: \textbf{Mean Average Precision} (MAP) and \textbf{Precision@1} (P@1).
MAP rewards methods that return suitable responses on the top and also rewards methods that return correct ranking of responses.
P@1 reports the fraction of suitable responses among the top 1 responses retrieved. All the results reported below are based on 5-fold cross-validation on the 422 queries.
We also perform a significance test using a paired \textbf{\textit{t}}-test with a significant level of 0.05.

We measure the performance of the logistic regression classifier for learning topic words in terms of \textbf{Accuracy} based on 5-fold cross-validation on the 2008 labeled words.

\subsubsection{Parameter Settings}
There are five parameters to be set in our experiments. We tune the best parameters with 5-fold cross-validation. Finally, we set $\alpha=0.8$ as the Jelinek-Mercer smoothing factor in Equation (\ref{equ:alpha}), $\beta=0.9$ to interpolate the post model and the response model in Equation (\ref{equ:beta_gamma}), $\gamma=0.5$ to interpolate the unigram language model and the translation model in Equation (\ref{equ:beta_gamma}).
We set $c=0$ in Equation (\ref{equ:logistic_regression}).
When training the weighs of features in Equation (\ref{equ:rankingsvm_linear}) with linear RankingSVM \citep{Herbrich:1999:RankingSVM}, we use a fixed penalty parameter (i.e., 50), as the performance is fairly insensitive to the choice of the parameter.

\subsection{Results of Basic Linear Matching Models} \label{subsec:results_basic}
We first evaluate the performance of the three basic linear matching models combined with the simple matching features in the learning to rank framework with Equation (\ref{equ:rankingsvm_linear}).

The results are shown in Table \ref{tab:results_basic}.
In the table, Q2R stands for the features based on the query-response similarity (Section \ref{subsubsec:Q2R_sim}) and query-response related simple matching features (Section \ref{subsec:simple_matching_features}).
Q2P stands for the features based on the query-post similarity (Section \ref{subsubsec:Q2P_sim}) and query-post related simple matching features (Section \ref{subsec:simple_matching_features}).
LatentMatch stands for the latent matching feature introduced in Section \ref{subsubsec:semantic_matching}.
As shown in the table, combining the three basic linear matching models with all the simple matching features achieves fairly good performance (row 4) and we name this model as Baseline, which will be used in the following subsections. In particular, we find that the LatentMatch feature helps slightly improve the overall performance on P@1.

%%\begin{table*}[!htb]\small
\begin{table*}\small
\centering
\caption{Comparison of different choices of features, where Q2R stands for the features based on the query-response similarity and query-response related simple matching features, Q2P stands for the features based on the query-post similarity and query-post related simple matching features, and LatentMatch stands for the latent matching feature. Row 4 combines the three basic linear matching models with all the simple matching features.} \label{tab:results_basic}
\begin{tabular}{|c|l|c|c|}
\hline

\textbf{\#} & \textbf{Model} & \textbf{MAP} & \textbf{P@1} \\ \hline \hline

1 & Q2R & 0.565 & 0.489 \\ \hline
2 & Q2R+Q2P & 0.621 & 0.567 \\ \hline
3 & Q2R+LatentMatch & 0.575 & 0.513 \\ \hline
4 & Q2R+Q2P+LatentMatch (Baseline) & 0.621 & 0.574 \\ \hline

\end{tabular}
\end{table*}

\subsection{Results of Combining all the Features} \label{subsec:results_combining_all}
Then, we further incorporate TransLM, DeepMatch and TopicWord as matching features into the learning to rank framework with Equation (\ref{equ:rankingsvm_linear}).

Table \ref{tab:results_combine_features} shows the comparison of different combinations of the matching features for retrieval-based STC.
Baseline stands for the model, which combines the three basic linear matching models with all the simple matching features, based on the learning to rank framework (see row 4 in Table \ref{tab:results_basic}).
From the table, we can see that the three new matching features can significantly improve the retrieval performance. When combining all the three new features with Baseline, the model achieves the best performance, which outperforms Baseline by 3.3 percent and 6.3 percent in terms of MAP and P@1, respectively (row 12 vs. row 5).

%%\begin{table*}[!htb]\small
\begin{table*}\small
\centering
\caption{Comparison of different combinations of the matching features for retrieval-based STC. \%impr\_MAP and \%impr\_P@1 stand for the percentage of improvements of the current model over the Baseline in terms of MAP and P@1, respectively.} \label{tab:results_combine_features}
%\addtolength{\tabcolsep}{-1pt}
\begin{tabular}{|c|l|c|c|c|c|}
\hline

\textbf{\#} & \textbf{Model} & \textbf{MAP} & \textbf{P@1} & \textbf{\%impr\_MAP} & \textbf{\%impr\_P@1} \\ \hline \hline

5 & Baseline & 0.621 & 0.574 & --- & --- \\ \hline \hline

6 & Baseline+TransLM & 0.641 & 0.605 & 2 & 3.1 \\ \hline
7 & Baseline+DeepMatch & 0.628 & 0.587 & 0.7 & 1.3 \\ \hline
8 & Baseline+TopicWord & 0.635 & 0.586 & 1.4 & 1.2 \\ \hline \hline

9 & Baseline+TransLM+DeepMatch & 0.643 & 0.625 & 2.2 & 5.1 \\ \hline
10 & Baseline+TransLM+TopicWord & 0.653 & 0.623 & 3.2 & 4.9 \\ \hline
11 & Baseline+DeepMatch+TopicWord & 0.641 & 0.606 & 2 & 3.2 \\ \hline \hline

12 & Baseline+TransLM+DeepMatch+TopicWord & \textbf{0.654} & \textbf{0.637} & \textbf{3.3} & \textbf{6.3} \\ \hline

\end{tabular}
\end{table*}

In order to clearly see the contributions of the three new features, we make comparisons between the model with and without each of the features.
Table \ref{tab:contribution_TransLM}, \ref{tab:contribution_DeepMatch} and \ref{tab:contribution_TopicWord} show the contributions of TransLM, DeepMatch and TopicWord, respectively.
Take Table \ref{tab:contribution_TransLM} as example, X and X+TransLM stands for the models without and with TransLM, respectively. X includes Baseline, Baseline+DeepMatch, Baseline+TopicWord, and Baseline+DeepMatch +TopicWord. \%impr means the percentage of improvements of X+TransLM over X in terms of MAP and P@1. Table \ref{tab:contribution_DeepMatch} and \ref{tab:contribution_TopicWord} are similar to Table \ref{tab:contribution_TransLM}. From the three tables, we have the following findings:

\begin{itemize}
\item The feature TransLM can bring at least 1.3 and 3.1 percent improvements in terms of MAP and P@1, respectively.
\item The feature DeepMatch can bring at least 1.3 percent improvements in terms of P@1, although it brings no much improvement in terms of MAP.
\item The feature TopicWord can bring at least 1.1 and 1.2 percent improvements in terms of MAP and P@1, respectively.
\end{itemize}

From the above analysis, we find that all the three features make significant contributions to the enhancement of performance. In addition, TransLM contributes the most, TopicWord the second largest, and DeepMatch the least.

\begin{table*}[!htb]\small
%\begin{table*}\small
\centering
\caption{Contributions of TransLM as a matching feature in the learning to rank framework for retrieval-based STC. X and X+TransLM stands for the models without and with TransLM, respectively. X includes Baseline, Baseline+DeepMatch, Baseline+TopicWord, and Baseline+DeepMatch+TopicWord. \%impr means the percentage of improvements of X+TransLM over X in terms of MAP and P@1.} \label{tab:contribution_TransLM}
%\addtolength{\tabcolsep}{-5.3pt}
\begin{tabular}{|c|c|c|c|c|c|}
\hline

 & \textbf{Model} & Baseline & Baseline+DeepMatch & Baseline+TopicWord & Baseline+DeepMatch+TopicWord \\ \hline \hline

\multirow{3}{*}{\textbf{MAP}} & X & 0.621 & 0.628 & 0.635 & 0.641 \\ \cline{2-6}
 & X\textbf{+TransLM} & 0.641 & 0.643 & 0.653 & 0.654 \\ \cline{2-6}
 & \%impr & 2 & 1.5 & 1.8 & 1.3 \\ \hline \hline

\multirow{3}{*}{\textbf{P@1}} & X & 0.574 & 0.587 & 0.586 & 0.606 \\ \cline{2-6}
 & X\textbf{+TransLM} & 0.605 & 0.625 & 0.623 & 0.637 \\ \cline{2-6}
 & \%impr & 3.1 & 3.8 & 3.7 & 3.1 \\ \hline

\end{tabular}
\end{table*}

\begin{table*}[!htb]\small
%\begin{table*}\small
\centering
\caption{Contributions of DeepMatch as a matching feature in the learning to rank framework for retrieval-based STC. X and X+DeepMatch stands for the models without and with DeepMatch, respectively. X includes Baseline, Baseline+TransLM, Baseline+TopicWord, and Baseline+TransLM+TopicWord. \%impr means the percentage of improvements of X+DeepMatch over X in terms of MAP and P@1.} \label{tab:contribution_DeepMatch}
%\addtolength{\tabcolsep}{-4.6pt}
\begin{tabular}{|c|c|c|c|c|c|}
\hline

 & \textbf{Model} & Baseline & Baseline+TransLM & Baseline+TopicWord & Baseline+TransLM+TopicWord \\ \hline \hline

\multirow{3}{*}{\textbf{MAP}} & X & 0.621 & 0.641 & 0.635 & 0.653 \\ \cline{2-6}
 & X\textbf{+DeepMatch} & 0.628 & 0.643 & 0.641 & 0.654 \\ \cline{2-6}
 & \%impr & 0.7 & 0.2 & 0.6 & 0.1 \\ \hline \hline

\multirow{3}{*}{\textbf{P@1}} & X & 0.574 & 0.605 & 0.586 & 0.623 \\ \cline{2-6}
 & X\textbf{+DeepMatch} & 0.587 & 0.625 & 0.606 & 0.637 \\ \cline{2-6}
 & \%impr & 1.3 & 2 & 2 & 1.4 \\ \hline

\end{tabular}
\end{table*}

\begin{table*}[!htb]\small
%\begin{table*}\small
\centering
\caption{Contributions of TopicWord as a matching feature in the learning to rank framework for retrieval-based STC. X and X+TopicWord stands for the models without and with TopicWord, respectively. X includes Baseline, Baseline+TransLM, Baseline+DeepMatch, and Baseline+TransLM+DeepMatch. \%impr means the percentage of improvements of X+TopicWord over X in terms of MAP and P@1.} \label{tab:contribution_TopicWord}
%\addtolength{\tabcolsep}{-4.8pt}
\begin{tabular}{|c|c|c|c|c|c|}
\hline

 & \textbf{Model} & Baseline & Baseline+TransLM & Baseline+DeepMatch & Baseline+TransLM+DeepMatch \\ \hline \hline

\multirow{3}{*}{\textbf{MAP}} & X & 0.621 & 0.641 & 0.628 & 0.643 \\ \cline{2-6}
 & X\textbf{+TopicWord} & 0.635 & 0.653 & 0.641 & 0.654 \\ \cline{2-6}
 & \%impr & 1.4 & 1.2 & 1.3 & 1.1 \\ \hline \hline

\multirow{3}{*}{\textbf{P@1}} & X & 0.574 & 0.605 & 0.587 & 0.625 \\ \cline{2-6}
 & X\textbf{+TopicWord} & 0.586 & 0.623 & 0.606 & 0.637 \\ \cline{2-6}
 & \%impr & 1.2 & 1.8 & 1.9 & 1.2 \\ \hline

\end{tabular}
\end{table*}

\section{Case Study} \label{sec:case_study}

To get a better understanding of the effectiveness of the matching features, we conduct case study of the features. We illustrate the results through several examples.
Section \ref{subsec:case study for Basic} shows the effectiveness of the basic linear matching features.
Sections \ref{subsec:case study for TransLM}, \ref{subsec:case study for DeepMatch}, and \ref{subsec:case study for TopicWord} show the effectiveness of using translation-based language model, deep matching model and topic-word model.
Section \ref{subsec:case study for FutureWork} gives several examples which are not addressed well with our current model and we will leave them to future work.

\subsection{The Effectiveness of Basic Linear Matching} \label{subsec:case study for Basic}

The basic linear matching features are mostly vector-space based, which are fairly good at capturing semantic relevance, as illustrated in Table \ref{tab:Example_effect_Basic_VSM}. The suitable responses are retrieved mainly because they have common words with the queries.
The experiments also show that we may find interesting and suitable responses that have no common words with the query, as show in the example in Table \ref{tab:Example_effect_LatentMatch}.
\comment{The latent space model however performs relatively poorly on long posts, where the topics of the texts cannot be well captured by the sum of the latent vectors of words.}

\begin{table*}[!htb]\small
%%\begin{table*}\small
  \centering
  \caption{Two illustrative examples of the effectiveness of basic matching models.}\label{tab:Example_effect_Basic_VSM}

    %\addtolength{\tabcolsep}{-4pt}
    \begin{tabular}{|l|p{12.5cm}|}
    \hline

    Query 1 & \tabincell{p{13cm}}{这是 西班牙 的 一 个 拥有 500 人口 的 小城镇 ， 你 猜 怎么着 ， 他们 甚至 有 一 个 赌场 ！  \\ It is a small town in Spain with 500 population, and guess what, they even have a casino!} \\ \hline

    Suitable Response 1 & \tabincell{p{12.5cm}}{如果 你 到 西班牙 旅游 ， 你 需要 花 一些 时间 在 那里。 \\ If you travel to Spain, you need to spend some time there.} \\ \hline \hline

    Query 2 & \tabincell{p{12.5cm}}{引自本杰明·富兰克林的一个名言：“我们都是天生的无知，但必须努力保持着愚蠢。” \\ One quote from Benjamin Franklin: ``We are all born ignorant, but one must work hard to remain stupid.''} \\ \hline

    Suitable Response 2 & \tabincell{p{12.5cm}}{本杰明·富兰克林是位智者，美国的开国元勋之一。 \\ Benjamin Franklin is a wise man, and one of the founding fathers of USA.} \\ \hline

    \end{tabular}

\end{table*}

\begin{table*}[!htb]\small
%%\begin{table*}\small
  \centering
  \caption{An illustrative example of the effectiveness of latent space matching model.}\label{tab:Example_effect_LatentMatch}

    \begin{tabular}{|l|p{6.5cm}|}
    \hline

    Query & \tabincell{p{6.5cm}}{英格兰 禁区 里 老 是 八 个 人 … … \\ Eight England players stand in the penalty area ...} \\
    \hline \hline

    Suitable Response 1 & \tabincell{p{6.5cm}}{那 场 比赛 真 经典 。 \\ What a classic match.} \\ \hline

    Suitable Response 2 & \tabincell{p{6.5cm}}{哈哈哈 仍然 是 1：0 。 还 没 看到 进球 。 \\ Hahaha, it is still 1:0, no goal so far.} \\ \hline

    \end{tabular}

\end{table*}

\subsection{The Effectiveness of Translation-based Language Model} \label{subsec:case study for TransLM}

The experimental results show that TransLM has superior performance when used as a feature in the learning to rank framework. Candidate post-comment pairs that do not share many common words with the query tend to be ranked low by the other matching models. However, the translation-based model is able to fill the lexical gap and find lexically dissimilar but semantically similar post-comment pairs, and rank them high. Table \ref{tab:Example_effect_TransLM} gives some retrieved post-comment pairs for a given query. We can see that the model with TransLM as one of the features can rank the suitable responses higher than that without TransLM. This is due to the word translation probabilities: $T$(夕阳$|$日落)=0.018, $T$(夕阳$|$黄昏)=0.012, and $T$(西西里岛$|$欧洲)=0.001.

\begin{table*}[!t]\small
%%\begin{table*}\small
  \centering
  \caption{An illustrative example of the effectiveness of translation-based language model. The ``Labels'' column indicates whether the candidate responses are suitable to the query. The ``Candidate Responses'' and ``Original Posts'' columns list the candidate responses with their original posts, respectively. The ``Before'' and ``After'' columns show the ranking of responses in the top 30 candidates before and after using TransLM as one of the features, respectively.}\label{tab:Example_effect_TransLM}

    \addtolength{\tabcolsep}{-4pt}
    \begin{tabular}{|l|p{8cm}|p{4cm}|c|c|}
    \hline

    \textbf{Query} & \multicolumn{4}{p{12cm}|}{\tabincell{p{12cm}}{小舟 划向 夕阳 （ 摄于 西西里岛 附近 ） 。 晚安 ！ \\ A boat rowed off into the sunset (taken in the vicinity of Sicily). Good Night!}} \\
    \hline \hline

    \textbf{Labels} & \textbf{Candidate Responses} & \textbf{Original Posts} & \textbf{Before} & \textbf{After} \\ \hline

    Suitable & \tabincell{p{7.5cm}}{真 不错 , 夕阳 西下 , 晚安 ! \\ Very beautiful sunset. Good night!} & \tabincell{p{4.5cm}}{西西里岛 的 日落 。 大家 晚安 ！ \\ Sicily's sunset. Good night!} & 1 & 2 \\ \hline

    Suitable & \tabincell{p{7.5cm}}{西西里岛 的 日落 。 大家 晚安 ！ \\ Sicily's sunset. Good night!} & \tabincell{p{4.5cm}}{西西里岛 的 日落 。 大家 晚安 ！ \\ Sicily's sunset. Good night!} & 2 & 1 \\ \hline

    Suitable & \tabincell{p{8cm}}{日落 夜 不 黑 的 欧洲 夜晚 \\ Sunset night, not black night in Europe} & \tabincell{p{4cm}}{苏黎世 日落 。 大家 晚安 ！ \\ Zurich sunset. Good night!} & 22 & 3 \\ \hline

    Suitable & \tabincell{p{8cm}}{种么 美 ， 想起 《 爱 在 黄昏 日落 时 》 的 场景 \\ What a beauty! I think of the ``Love in the Sunset'' scene} & \tabincell{p{4cm}}{巴黎 日落 。 大家 晚安 ！ \\ Paris sunset. Good night!} & 25 & 4 \\ \hline

    \end{tabular}

\end{table*}

\subsection{The Effectiveness of Deep Matching Model} \label{subsec:case study for DeepMatch}

Table \ref{tab:Example_effect_DeepMatch} shows some retrieved responses for a given query. From the table, we can see that although the two suitable responses share almost no common words with the query, the model with DeepMatch as one of the features can match them well and rank them higher than that without the feature.

\begin{table*}[!htb]\small
%%\begin{table*}\small
  \centering
  \caption{An illustrative example of the effectiveness of deep matching model. The ``Labels'' column indicates whether the candidate responses are suitable to the query. The ``Candidate Responses'' column lists the candidate responses. The ``Before'' and ``After'' columns show the ranking of responses in the top 30 candidates before and after using DeepMatch as one of the features, respectively.}\label{tab:Example_effect_DeepMatch}

    \addtolength{\tabcolsep}{-4pt}
    \begin{tabular}{|l|p{10cm}|c|c|} \hline

    \textbf{Query} & \multicolumn{3}{p{12cm}|}{\tabincell{p{12cm}}{洗车 过程 中 就 开始 下雨 ， 这 是 一 种 什么样 的 衰 ! \\ It began to rain when I was washing the car. Such a bad luck!}} \\ \hline \hline

    \textbf{Labels} & \textbf{Candidate Responses} & \textbf{Before} & \textbf{After} \\ \hline

    Suitable & \tabincell{p{10cm}}{经常 听 朋友 这么 说 。 看来 还 是 不 洗 的 划算 ， 哈哈 ！ \\ I often hear that from my friends. It is better not to wash it to save money, haha!} & 3 & 2 \\ \hline

    Suitable & \tabincell{p{10cm}}{同 命人 哇 ， 你 俩 互相 安慰 安慰 \\ You guys have the same situation, so at least you can comfort each other.} & 6 & 1 \\ \hline

    \end{tabular}

\end{table*}

\subsection{The Effectiveness of Topic-Word Model} \label{subsec:case study for TopicWord}

Table \ref{tab:Example_effect_TopicWord_1} gives some retrieved responses for a given query. From the table, we can clearly see that the unsuitable responses, which do not share topic words with the query, are ranked lower when using TopicWord as one of the features; while the suitable response, which share topic words with the query, is ranked higher when using TopicWord as one of the features.
More specifically, the word ``菜鸟(rookie)'' in the query is not a topic word, thus has a low term weight (i.e., the probability of being a topic word). Although the word ``菜鸟(rookie)'' in the unsuitable responses is a topic word with a high term weight, the cosine similarities between the query and the two unsuitable responses are still not high after using the term weighting of topic words. Moreover, the suitable response is ranked higher mainly because it has common topic words ``控制(control), 工具(tool)'', with the query, which are assigned higher weights.

\begin{table*}[!t]\small
%\begin{table*}\small
  \centering
  \caption{An illustrative example of the effectiveness of topic-word model. The ``Labels'' column indicates whether the candidate responses are suitable to the query. The ``Candidate Responses'' column lists the candidate responses. The ``Before'' and ``After'' columns show the ranking of responses in the top 30 candidates before and after using TopicWord as one of the features, respectively.}\label{tab:Example_effect_TopicWord_1}

    \addtolength{\tabcolsep}{-4pt}
    \begin{tabular}{|l|p{12.2cm}|c|c|} \hline

    \textbf{Query} & \multicolumn{3}{p{14.1cm}|}{\tabincell{p{14.1cm}}{选用 \textbf{代码 控制 工具} 的 时候 ， 还是 用 \textbf{SVN} ， 不 要 用 \textbf{GIT} ， 虽然 \textbf{GIT} 比 \textbf{SVN} 强大 很多 。 因为 如果 团队 里 有 一 个 菜鸟 ， 他 各种各样 关于 \textbf{GIT} 的 问题 会 烦 死 你 。 \textbf{SVN} 的 优势 就 是 极其 简单 ， 菜鸟 也 可以 很 快 掌握 \\ When choosing the \textbf{code control tool}, use \textbf{SVN} instead of \textbf{GIT}, although \textbf{GIT} is more powerful than \textbf{SVN}. The reason is that if a team has a rookie, he will bother you with all sorts of questions about \textbf{GIT}. The advantage of \textbf{SVN} is it is so simple that a rookie can grasp it quickly enough.}} \\ \hline \hline

    \textbf{Labels} & \textbf{Candidate Responses} & \textbf{Before} & \textbf{After} \\ \hline

    Unsuitable & \tabincell{p{12.2cm}}{我 是 \textbf{保密员} 、 \textbf{菜鸟} 与 \textbf{90后} \textbf{努力} 为了 不 成为 老 \textbf{菜鸟} \\ I am a \textbf{confidential staff}, \textbf{rookie} and \textbf{the generation after 90s}, \textbf{making efforts} in order not to become an old \textbf{rookie}. } & 1 & 9 \\ \hline

    Unsuitable & \tabincell{p{12.2cm}}{我 ， 还 有 很多 用户 。 都 是 \textbf{菜鸟} 。 \\ I and also a lot of users are \textbf{rookies}.} & 2 & 4 \\ \hline

    Suitable & \tabincell{p{12.2cm}}{一 个 \textbf{版本 控制} 的 \textbf{工具} \\ A \textbf{version control tool}} & 9 & 1 \\ \hline

    \end{tabular}

\end{table*}

\comment{
\begin{table*}[!htb]\small
%\begin{table*}\small
  \centering
  \caption{Another illustrative example of the effectiveness of topic-word model. The ``Labels'' column indicates whether the candidate responses are suitable to the query. The ``Candidate Responses'' column lists the candidate responses. The ``Before'' and ``After'' columns show the ranking of responses in the top 30 candidates before and after using TopicWord as one of the features, respectively.}\label{tab:Example_effect_TopicWord_2}

    \addtolength{\tabcolsep}{-4pt}
    \begin{tabular}{|l|p{12.2cm}|c|c|} \hline

    \textbf{Query} & \multicolumn{3}{p{14.1cm}|}{\tabincell{p{14.1cm}}{【 \textbf{日本 美食} 】 “ \textbf{油そば} （ \textbf{abura}-\textbf{soba} ) ” ， 据说 是 发源于 \textbf{日本} 多摩 地区 的 \textbf{面条} 。 先 在 面碗 里 倒入 一 丁点 加入 酱油 和 猪油 的 调料 ， 然后 依照 个人 口味 放入 辣 油 、 醋 等 调料 ， 和 煮好 的 面条拌 在一起 。 最后 配 以 香葱 、 笋干 、 叉烧 肉 、 怎 一 个 香字 了 得 。 \\ {[\textbf{Japanese food}]}\textbf{abura-soba}, is said to be originated in the Tama area \textbf{noodles} in \textbf{Japan}. First put a little bit of soy sauce and spices lard into a bowl, and then mix vinegar, spices, and cooked noodles together according to personal taste. Finally add some scallions, bamboo shoots, pork meat. How a word of incense is.}} \\ \hline \hline

    \textbf{Labels} & \textbf{Candidate Responses} & \textbf{Before} & \textbf{After} \\ \hline

    Unsuitable & \tabincell{p{12.2cm}}{建议 死 打 。 。 。 还 是 吃 \textbf{猪油} 和 \textbf{菜油} 吧 \\ Suggest death play... or eat \textbf{lard oil} and \textbf{vegetable oil}} & 1 & 16 \\ \hline

    Unsuitable & \tabincell{p{12.2cm}}{\textbf{配方} 就 是 很 \textbf{味千} 啊 。 但是 要 \textbf{看料} 。 。 如果 是 真的 猪骨 和 鱼干 熬 汤 就 一定 会 好吃 ， 但 \textbf{味千} 是 \textbf{味精} 。 \\ The \textbf{recipe} is \textbf{Ajisen}, but \textbf{depends on the material}.. If it is true pig bone and dried fish soup, it will be certainly delicious, but \textbf{Ajisen} is \textbf{MSG}.} & 2 & 8 \\ \hline

    Suitable & \tabincell{p{12.2cm}}{去 \textbf{日本} 的 时候 \textbf{吃} 过 ， 没有 想象 中 的 那 没 \textbf{好吃} \\ I have \textbf{eat} this when I went to \textbf{Japan}, it is not \textbf{delicious} as imagined.} & 7 & 5 \\ \hline

    Suitable & \tabincell{p{12.2cm}}{為甚 麼變 成 了 \textbf{日本 美食} \\ Why it is turned into \textbf{Japanese food}} & 25 & 1 \\ \hline

    Suitable & \tabincell{p{12.2cm}}{\textbf{日本 美食} 就 是 诱人 啊 \\ \textbf{Japanese food} is temping} & 26 & 2 \\ \hline

    \end{tabular}

\end{table*}
}

\subsection{Some Failed Issues} \label{subsec:case study for FutureWork}
In this subsection, we show several issues which cannot be addressed well with our current model and we will leave them to future work.

\subsubsection{Entity Association}

Entity association is only partially addressed with features like query-response cosine similarity, with entity names treated as words, which is apparently not enough for preventing the following type of mistakes (see Table \ref{tab:Example_failed_entity_association}) when the candidate response and the query match well on other parts. Actually, for query 1, we only need to modify the word ``李教授(Prof. Li)'' in the unsuitable response 1 to the word ``王教授(Prof. Wang)'', to get expected response 1, which is suitable to the query 1. For query 2, we only need to modify the word ``画(drawing)'' in the unsuitable response 2 to the word ``瓷器(china)'', to get expected response 2, which is suitable to the query 2.

\begin{table*}[!htb]\small
%%\begin{table*}\small
  \centering
  \caption{Two failed examples on entity association.}\label{tab:Example_failed_entity_association}

    %\addtolength{\tabcolsep}{-4pt}
    \begin{tabular}{|l|p{11.5cm}|}
    \hline

    Query 1 & \tabincell{p{13cm}}{\textbf{王教授}将会在下学期开一门自然语言处理的课程。 \\ \textbf{Professor Wang} will give a course on natural language processing, starting next semester.} \\ \hline

    Unsuitable Response 1 & \tabincell{p{11.5cm}}{羡慕啊。。我希望在未来的某段时间里我也可以参加\textbf{李教授}的课程。 \\ Jealous.. I wish I can attend \textbf{Prof. Li}'s course too some time in the future.} \\ \hline

    Expected Response 1 & \tabincell{p{11.5cm}}{羡慕啊。。我希望在未来的某段时间里我也可以参加\textbf{王教授}的课程。 \\ Jealous.. I wish I can attend \textbf{Prof. Wang}'s course too some time in the future.} \\ \hline \hline

    Query 2 & \tabincell{p{11.5cm}}{台北故宫博物院精美\textbf{瓷器}展 \\ The fine \textbf{china} from Exhibition at the National Palace Museum in Taipei} \\ \hline

    Unsuitable Response 2 & \tabincell{p{11.5cm}}{这\textbf{画}看起来很不错。台北故宫博物院到处是国宝。 \\ This \textbf{drawing} looks so nice. National Palace Museum in Taipei is full of national treasures.} \\ \hline

    Expected Response 2 & \tabincell{p{11.5cm}}{这\textbf{瓷器}看起来很不错。台北故宫博物院到处是国宝。 \\ This \textbf{china} looks so nice. National Palace Museum in Taipei is full of national treasures.} \\ \hline

    \end{tabular}

\end{table*}

\subsubsection{Logic Consistency}

Our current model does not directly maintain the logic consistency between the candidate response and the query, since logic consistency requires a deeper analysis of the texts, and therefore hard to implement. Table \ref{tab:Example_failed_logic_consistency} shows two examples which are semantically relevant, and correct with respect to speech act, but logically inappropriate.

\begin{table*}[!htb]\small
%%\begin{table*}\small
  \centering
  \caption{Two failed examples on logic consistency.}\label{tab:Example_failed_logic_consistency}

    %\addtolength{\tabcolsep}{-4pt}
    \begin{tabular}{|l|p{13cm}|}
    \hline

    Query 1 & \tabincell{p{13cm}}{查 了 一下 ， 王凤仪 先生 应该 不 是 我 的 外祖父 ， 虽然 他们 的 事迹 相似 ， 并且 同 称 「 王 大 善 人 」 。
 \\ I checked. Wang Fengyi is not my maternal grandfather, although they've done similar deeds and both are called ``Wang the Well-doer''.} \\ \hline

    Unsuitable Response 1 & \tabincell{p{12.5cm}}{您 外祖父 是 王凤仪 先生 啊
 \\ Wow, Wang Fengyi is your maternal grandfather.} \\ \hline \hline

    Query 2 & \tabincell{p{12.5cm}}{全球 最 薄 最 紧凑 大 屏旗舰 智能 手机 Ascend P1s 首发 是 UMTS /HSPA + 版本 ， 还 会 有 中国 移动 TD 版本 。
 \\ The debut of the world's thinnest, most compact, big-screen flagship smartphone Ascend P1s is UMTS / HSPA + version, there will be China Mobile TD version.} \\ \hline

    Unsuitable Response 2 & \tabincell{p{12.5cm}}{有 没 有 移动 TD 版本 的 ？ \\ Will there be any version of China mobile TD version?} \\ \hline

    \end{tabular}

\end{table*}

\section{Conclusions} \label{sec:conclusions}

In this paper we have proposed a retrieval-based model for short text conversation (STC), to leverage massive data collected from social media. Our experiments show that the retrieval-based model performs reasonably well, when combined with a set of carefully designed matching features and a huge repository of conversation data.

This work opens to several interesting directions for future work with regard to STC.
When performing retrieval-based STC, we need to consider matching between query and response in terms of semantic relevance. In addition, we may also need to consider matching between query and response in terms of speech act, sentiment, entity association, logic consistency and discourse structure. How to model these factors and how to enhance the accuracy based on the factors in STC are open and challenging issues.

%\section*{Acknowledgement}

%This work is supported in part by China National 973 project 2014CB340301.



\section*{References}
\bibliographystyle{elsarticle-harv}
\bibliography{library}


%% \begin{thebibliography}{00}


%% \bibitem[ ()]{}

%% \end{thebibliography}

\clearpage\end{CJK*}
\end{document}